\documentclass[a4paper]{panl}
\usepackage{seqsplit}
\usepackage{cite}
\usepackage{wrapfig}
\usepackage{graphicx}
\usepackage{amssymb}
\usepackage{amsfonts}
\usepackage{amsmath}
\usepackage{longtable}
\usepackage{rotating}
\usepackage{lscape}
\usepackage{epsfig}
\usepackage{multirow}

\def\bbljan{Jan}
\def\bblfeb{Feb}
\def\bblmar{Mar}
\def\bblapr{Apr}
\def\bblmay{May}
\def\bbljun{Jun}
\def\bbljul{Jul}
\def\bblaug{Aug}
\def\bblsep{Sep}
\def\bbloct{Oct}
\def\bblnov{Nov}
\def\bbldec{Dec}

\def\apj{ApJ}
\def\apjl{ApJL}
\def\apjs{ApJ Suppl. Ser.}

\def\jcap{JCAP}
\def\mnras{MNRAS}
\def\nat{Nature}

\def\sovast{Sov. Astron}

\def\prd{Phys. Rev. D}

\def\M{{\cal M}}
\def\R{{\cal R}}
\def\D{{\cal D}}
\def\BibDash{--}

\def\beq#1{\begin{equation}\label{#1}}
\def\eeq{\end{equation}}
\def\beqa#1{\begin{eqnarray}\label{#1}}
\def\eeqa{\end{eqnarray}}

\def\eqn#1{~(\ref{#1})}
\def\myfrac#1#2{\left(\frac{#1}{#2}\right)}

\def\comment#1{\relax}

\originalTeX
\begin{document}
% Journal sections (see http://pkp.jinr.ru/index.php/PEPAN_LETTERS/about/editorialPolicies#focusAndScope)
%\issuearea{Physics of Elementary Particles and Atomic Nuclei. Theory}
% or in Russian
%\issuearea{ФИЗИКА ЭЛЕМЕНТАРНЫХ ЧАСТИЦ И АТОМНОГО ЯДРА. ТЕОРИЯ}

\title{On the primordial binary black hole mergings in LIGO-Virgo-Kagra data}
\maketitle
%\authors{K.A.\,Postnov$^{a,b,}$\footnote{E-mail: first.author@email.ru},S.\,Author$^{b,}$\footnote{E-mail: second.author@email.ru}}
\setcounter{footnote}{0}
\authors{K.A.\,Postnov$^{a,}$\footnote{E-mail: pk@sai.msu.ru},
N.A.\,Mitichkin$^{a,}$\footnote{E-mail: mitichkin.nikita99@mail.ru}}
%\from{$^{a}$\,Affiliation 1}
\from{$^{a}$\, Sternberg Astronomical Institute, M.V. Lomonosov Moscow State University, Universitetskij pr. 13, 119234, Moscow, Russia}
%\from{$^{b}$\, }
%\from{$^{b}$\,Место работы автора 2}

%%%%%%%%%%%%% line numbering
%\linenumbers
%%%%%%%%%%%%% line numbering

\begin{abstract}
We briefly discuss a possible cosmological implication of the observed binary black hole  mergings detected by LIGO-Virgo-Kagra collaboration (GWTC-3 catalogue) for the primordial black hole (PBH) formation in the early Universe. We show that the bumpy chirp mass distribution of the LVK BH+BH binaries can be fit with two distinct and almost equal populations: (1) astrophysical mergings from BH+BH formed in the modern Universe from evolution of massive binaries and (2) mergings of binary PBHs  with initial log-normal mass distribution. We find that the PBH central mass ($M_c\simeq 30 M_\odot$) and distribution width  derived from the observed LVK chirp masses are almost insensitive to the assumed double PBH formation model. To comply with the observed LVK BH+BH merging rate, the CDM PBH mass fraction should be $f_{pbh}\sim 10^{-3}$ but can be higher if PBH clustering is taken into account.  
\end{abstract}
\vspace*{6pt}

\label{sec:intro}
\section*{Introduction}

%Gravitational-wave astronomy is a new and rapidly developing field of science that investigates astrophysical sources of gravitational waves (GW) using methods of multi-wavelength ground-based and space astronomy (from radio to gamma-ray wavelengths). 
The first detection of merging of massive black holes (BHs) in the binary system GW150914 by the laser interferometers LIGO \cite{PhysRevLett.116.061102} heralded a new era in astronomical observations -- gravitational-wave (GW) astronomy.
%The merger of two BHs as the most likely astrophysical source in the frequency range of sensitivity of the LIGO GW interferometers $\sim 10-1000$~Hz was anticipated from the evolution of binary stars \cite{1993MNRAS.260..675T, 1997MNRAS.288..245L, 1997NewA....2...43L, 1997AstL...23..492L,2005PhyU...48.1235G}. 
A truly multimessenger astronomy started after the detection of the first coalescing binary neutron star (NS)  GW170817 \cite{PhysRevLett.119.161101}, which was associated with a short gamma-ray burst GRB170817A and accompanied by a follow-up multi-wavelength electromagnetic signal  \cite{2017ApJ...848L..12A} arising from thermal and non-thermal emission of matter during the merger (the so-called 'kilonova' \cite{2010MNRAS.406.2650M}). 
Since then, the LVK (LIGO-Virgo-Kagra) collaboration has completed three  observing runs O1-O3. The detected astrophysical sources are summarized in the GWTC-3 catalogue \cite{2021arXiv211103606T} and their properties are discussed in \cite{2021_GWTC3binaries}. The catalogue includes $\sim 90$  BH+BH, BH+NS and double NS coalescences detected with a signal-to-noise ratio $S/N>8$.  The analysis of these data and their interpretation  are reviewed, e.g., in \cite{2022Galax..10...76S,2022LRR....25....1M}. 

The number of detected sources enables statistical studies of  possible populations of coalescing compact binaries. Coalescing binary NS and BH+NS systems are thought to result from the evolution of massive binaries \cite{2022LRR....25....1M}. The coalescing binary BHs can be formed in different ways. The main astrophysical channels include the evolution of massive binary systems \cite{1993MNRAS.260..675T, 1997MNRAS.288..245L,2005PhyU...48.1235G,2016Natur.534..512B}, 
the dynamical evolution in dense stellar systems \cite{2016ApJ...824L...8R,2021ApJ...915L..35K} or 
the origin in dense gas environments in galactic nuclei \cite{2021ApJ...908..194T}.

The very first detection of the coalescing binary BH GW1250914 with unusually high masses of the components, $\sim 30 M_\odot$, much exceeding the
dynamically estimated masses of BHs in the Galactic BH X-ray binaries \cite{2016ApJS..222...15T}, revived the interest to the primordial BHs (PBHs) that can be formed in the early Universe at the radiation-dominated stage 
\cite{2016PhRvL.116t1301B,2016JCAP...11..036B,2016PhRvL.117f1101S}.
Indeed, in the standard PBH formation mechanism from primordial cosmological perturbations \cite{1967SvA....10..602Z,1974MNRAS.168..399C}, the mass of a PBH should be 
of the order of mass inside the cosmological horizon, 
$M_\mathrm{hor} \approx (2.2 \times 10^5 M_\odot)(t/[\mathrm{s}])$. For example, 
at the QCD phase transition epoch at the temperature $T\sim 100$~MeV, 
the mass inside the cosmological horizon $M_\mathrm{hor}^{\mathrm{QCD}} \approx 8 M_\odot (100 \,{\rm MeV}/ T_\mathrm{QCD})^2$ \cite{2020JCAP...04..052D}, suggestively close the BH masses estimated in the LVK sources. 
The formation of PBHs from cosmological perturbations and constraints on their abundance as possible CDM candidates are studied in many papers, e.g. \cite{2020ARNPS..70..355C,2021RPPh...84k6902C,2022PhRvD.106l3526F,2022arXiv220906196E}.
Therefore, one of the most intriguing cosmological implications of the GW astronomy results is the possibility that some of the detected binary BH coalescences can be PBHs formed in the early Universe at $t\sim 10^{-5}$~s after the Big Bang.

In this letter, we present the result of the analysis of the chirp mass distribution of the LVK binary BHs by a particular model including two populations of binary BHs -- with stellar masses $\sim 10 M_\odot$ formed from the evolution of massive stars in the modern Universe, and the PBHs with log-normal mass spectrum predicted by the modified Affleck-Dyne baryogenesis model \cite{1993PhRvD..47.4244D,2009NuPhB.807..229D}.

\section*{Chirp mass distribution of LVK binary black holes}

%\section{Binary BHs detected in GW observations}
From GW observations of a coalescing binary system with masses $m_1$ and $m_2$ and dimensionless angular momenta $a_i^*=J_ic/(Gm_i^2)$ ($G$, $c$ are the Newtonian gravitational constant and speed of light, respectively) it is possible to estimate (with varying accuracy depending on the signal strength) the following binary parameters: (1) the chirp-mass $\M=\frac{(m_1m_2)^{3/5}}{(m_1+m_2)^{1/5}}$ in the observer's system $\M_\mathrm{det}$\footnote[1]{Chirp-mass in the observer's frame is determined by measuring signal frequency during the inspiraling along quasi-Keplerian orbits before merging and is degenerate by source redshift $z$ $\M_\mathrm{det}=(1+z)\M$}, the individual component masses  and the binary mass ratio $q=m_2/m_1\le 1$, (2) the mass-weighted projection of individual spins on the orbital angular momentum (the effective spin) prior to the merging $\chi_{\mathrm{eff}}=\frac{m_1a_1^*\cos\theta_1+m_2a_2^*\cos\theta_2}{m_1+m_1}$, the projection of the total spin of the components on the orbital plane $\chi_p$, the spin of the resulting object after the merger $a^*_\mathrm{fin}$, (3) the luminosity distance to the source $D_l(z)$ (redshift $z$ in a given cosmological model) \cite{1986Natur.323..310S}. The chirp mass $\M$ is derived the most accurately. The effective spins of the LVK BH+BH mergings are less certain: they are close to zero with some outliers \cite{2021_GWTC3binaries}, however, their determination is model-dependent (especially for very massive coalescences, see, e.g., the discussion of the component spins in GW190521 in ref. \cite{2023NatAs...7...11G}). Therefore, here we will not address the problem of BH+BH effective spins.

%The classification of the merging binary system type by the GW signal is usually based on the estimation of the component masses -- a compact object with mass $m>3 M_\odot$ is known to be a BH \cite{1974PhRvL..32..324R,2016PhR...621..127L}{\footnote[2]{Maximal known mass of a NS in millisecond pulsar PSR J0952-0607 is $M_{PSR}\approx 2.35 M_\odot$ \cite{2022ApJ...934L..17R}.}. The source type  can be also identified from the GW signal alone by machine learning methods \cite{Qiu:2022wub}.

In Ref. \cite{2020JCAP...12..017D}, we have analyzed the chrip mass distribution constructed using  the GWTC-2 catalogue and open O3 LIGO-Virgo data. We have found that a population of PBHs with log-normal spectrum in the form proposed by \cite{1993PhRvD..47.4244D}
\beq{e:dNdM}
\frac{dn}{dm}=\mu^2\exp\left[-\gamma\ln^2\myfrac{m}{M_0}\right]
\eeq
($\mu$ is the normalization constant, $\gamma$ and $M_0$ are free parameters)\footnote[2]{After the mass renormalization $M_0=M_c\exp[-1/(2\gamma)]$ the PBH mass distribution can be written as 
$
F(m)=\sqrt{\frac{\gamma}{\pi}}\frac{1}{m}\exp\left[-\gamma\ln^2\myfrac{m}{M_c}\right]
$
such that $\int F(m) dm=1$ ($\gamma=1/2\sigma^2$ turns this expression into the standard log-normal form).
} can adequately fit the overall chirp mass distribution of the LVK sources for $M_0\sim 13-19 M_\odot$, $\gamma\sim 1$. Similar parameters of PBHs with log-normal mass spectrum were obtained from the analysis of LVK data in \cite{2019JCAP...02..018R,2022arXiv221016094L}. 

However, a single PBH population is expected to produce a smooth chirp-mass distribution of coalescing binaries, which is apparently not the case. There are two statistically significant bumps in the observed GWTC-3 chirp mass distribution of binary BHs at $\M\sim 10 M_\odot$ and $\M\sim 30 M_\odot$ \cite{2023arXiv230100834F}. The low-mass bump can be fitted with a population of binary BHs formed from massive stellar evolution. The high-mass bump requires a population of BHs with masses up to 100 $M_\odot$. 
BH masses $\sim 60 M_\odot-120 M_\odot$ are difficult to produce 
from the standard stellar evolution because of the pulsational pair instability due to electron-positron pair production in the stellar core, leading to a pair-instability supernovae explosion \cite{2017ApJ...836..244W,2021ApJ...912L..31W}.
Such BH masses in the evolution scenario of isolated massive binary systems can be obtained only under special assumptions \cite{2020ApJ...902L..36F}. Other possibilities include, for example, a hierarchical growth of BH masses in successive mergings in dense stellar systems mentioned above \cite{2021ApJ...915L..35K}, the formation of massive BHs from Population III stars with primordial elemental abundance \cite{2021ApJ...910...30T} and in the vicinity of active galactic nuclei \cite{2021ApJ...908..194T}. 

\begin{figure}[t]
\begin{center}
\includegraphics[width=127mm]{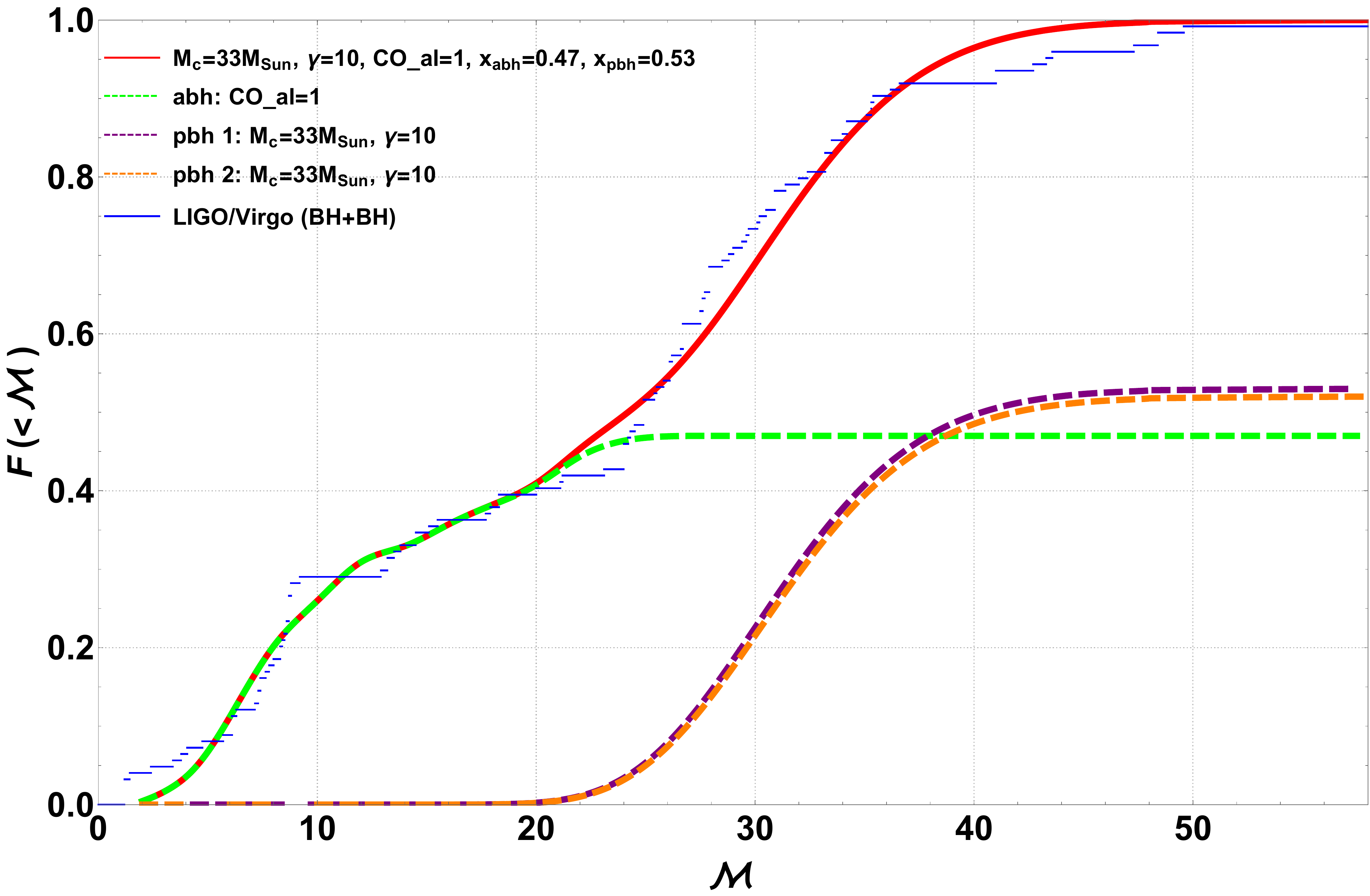}
\vspace{-3mm}
\caption{The observed (blue step-like curve) and model (red solid curve) distribution function of the chirp-masses of coalescing binary BHs from the LVK GWTC-3 catalogue \cite{2021arXiv211103606T}. The model includes almost equal contributions from coalescences of astrophysical binary BHs from ref. \cite{2019MNRAS.483.3288P} (green dashed curve) and primordial BHs with the initial log-normal mass spectrum \cite{1993PhRvD..47.4244D} (1) with parameters $M_c=33 M_\odot$, $\gamma=10$ (magenta and orange dashed curves for the PBH formation model 'pbh 1' \cite{1998PhRvD..58f3003I} and 'pbh 2' \cite{2019JCAP...02..018R}, respectively). Both models give the CDM PBH fraction $f_{pbh}\sim 10^{-3}$ for the observed BH+BH merging rate $0.5\times 25$ Gpc$^{-3}$~yr$^{-1}$ \cite{2021_GWTC3binaries}. The combined distribution (red) fits the empirical distribution function (blue) at 90\% CL according to the modified KS-test. \label{f:1} }
\end{center}
\vspace{-5mm}
\end{figure}

Thus, we assume that a fraction $x_{abh}$ of the LVK BH+BH mergings comes from the astrophysical channel (evolution of massive binaries). We use the population of binary BHs calculated in \cite{2019MNRAS.483.3288P} for different BH formation models. The expected chirp mass distribution functions $F(<\M)$ for these BH+BH populations with taking into account the evolution of star-formation rate and elemental abundance in galaxies with redshift $z$ is presented in Fig. 5 of ref. \cite{2020JCAP...12..017D}. We have found (the dashed green curve in Fig. \ref{f:1}) that the model of BH formation from the collapse of the CO-core of a massive star with the common envelope efficiency of the close binary evolution $\alpha_{CE}=1$ best fits the low-mass bump in the observed LVK chirp mass distribution for $x_{abh}=0.47$. 

The high-mass bump in the LVK GWTC-3 chirp masses can be fitted by a population of coalescing PBHs with the log-normal spectrum \eqn{e:dNdM}. To calculate the chirp mass distribution as detected by the GW interferometer with a given sensitivity, we have used the same method as in \cite{2020JCAP...12..017D}. In this approach, we have approximated the detection horizon with the S/N=8 sensitivity to coalescing binaries with chirp mass $\M$ as $
D_h(\mathcal M) = 122 \mathrm{Mpc} \left( \frac{\mathcal{M}}{1.2 \ M_\odot} \right)^{5/6}
$. Equating $D_h(\M)=d_l(z)$ ($d_l(z)$ is the luminosity distance in a given cosmological model) enables us 
to find the limiting redshift up to which the detector is sensitive for a given chirp mass $z_\mathrm{lim}(\M)$. 
From the differential merging rate $\R(z)$ (which is different in different PBH formation models, see below), we 
calculate the total PBH merging rate $\D\R(z_\mathrm{lim}(\M))$ as a function of $\M$ and the cumulative chirp mass distribution $F(<\M)$. 

There are several models of double PBH formation \cite{1997ApJ...487L.139N,2017PhRvD..96l3523A,2019JCAP...02..018R}. 
Here we will use the dynamical mechanism proposed in \cite{1997ApJ...487L.139N} and further elaborated in \cite{1998PhRvD..58f3003I}. In the latter paper, a 3-body interaction is involved to form a binary PBH. 
The merging rate per comoving volume is calculated as $\frac{d\R(z)}{dm_1dm_2}= n_{pbh}P(t(z))$, where $P(t(z))\propto t(z)^{-34/37}$ is a probability for a binary with $m_1$ and $m_2$ to merge at cosmic time $t(z)$ integrated over the 3d mass distribution $F(m_3)$ \cite{1998PhRvD..58f3003I}. Its application to the first LIGO event yielded $f_{pbh}\lesssim 10^{-3}$ \cite{2016PhRvL.117f1101S}. 
With the detection rate of LVK coalescing BH+BH binaries $\R\sim 16-61$ per Gpc$^3$ per year \cite{2021_GWTC3binaries}, the fraction $x_{pbh}=0.53 $ of the merging binary PBH means the binary PBH merging rate ${\cal R}\sim 8-30$ per Gpc$^3$ per year. This enables us to constrain the PBH fraction in cold dark matter (CDM) density $f_{pbh}$ for the assumed model of PBH formation and evolution, $f_{pbh}\approx 6\times 10^{-4}$, which is consistent with the previous constraints for this model \cite{2016PhRvL.117f1101S}. The best-fit parameters of the log-normal PBH distribution are $M_c\approx 30 M_\odot$, $\gamma\approx 10$ (the dashed magenta curve in Fig. \ref{f:1}). 

%============================= Fig. 1 ================================
%============================= Fig. 1 ================================

A different and widely used model of PBH formation \cite{2019JCAP...02..018R} gives the 
differential merging rate as a function  of redshift $z$ \cite{2022arXiv220906196E}
\[
\frac{d\R(z)}{dm_1dm_2}=\frac{3.2\times 10^6} {\mathrm{Gpc}^{3}\mathrm{yr}^{-1}} S(m_1,m_2,z) f_{pbh}^{\frac{53}{57}}F(m_1)F(m_2)\frac{m_1m_2}{\langle m\rangle^2}\myfrac{t(z)}{t_0}^{-\frac{34}{37}}M^{-\frac{32}{37}}\eta^{-\frac{34}{37}}
\]
($S\le 1$ is the suppression factor \cite{2019JCAP...02..018R,2021JCAP...03..068H} which we will set to 1 below, $t_0$ is the age of the Universe, $M=(m_1+m_2)/M_\odot$, $\eta=m_1m_2/(m_1+m_2)^2$, the mean mass $\langle m\rangle=\int mF(m)dm=e^\frac{1}{4\gamma}M_c$ for our choice of the log-normal mass function). 
Its application to the GWTC-2 LIGO-Virgo data yielded 
$f_{pbh}<0.002$ \cite{2021JCAP...03..068H} and suggested the existence of two populations of BH+BH binaries. Changing variables $(m_1,m_2)\to (\M,q)$ with the Jacobian 
$J(\M,q)=\frac{(1+q)^{2/5}}{q^{6/5}}\M$ yields 
\beq{}
\frac{d\R(z)}{d\M dq}\propto f_{pbh}^{\frac{53}{57}}F(\M)F(q)\frac{(1+q)^\frac{8}{5}}{q^{\frac{9}{5}}}\frac{\M^\frac{79}{37}}{M_c^2e^\frac{1}{2\gamma}}\myfrac{t(z)}{t_0}^{-\frac{34}{37}}\,.
\eeq
After substituting, as in \cite{2020JCAP...12..017D}, $z\to z_\mathrm{lim}(\M)$ and integrating over $q$, we calculate  the cumulative distribution 
\beq{}
\R(<\M)=\int\limits_0^\M\int\limits_0^{z_\mathrm{lim}(\M)}\int\limits_0^1\frac{d\R}{d\M dq}\frac{dV}{dz}\frac{dz}{1+z}dq,\quad F(<\M)=\frac{\R(<\M)}{\R(<\infty)}
\eeq
and the total PBH merging rate $\R(<\infty)$ in this model. It is not very surprising that the best-fit values of the initial log-normal PBH mass function in this case are close to those in the model \cite{1998PhRvD..58f3003I}, $M_c\approx 30 M_\odot$, $\gamma\approx 10$, and the CDM PBH fraction is $f_{PBH}\simeq 10^{-3}$. This is because for the log-normal initial PBH mass functions $F(m_1), F(m_2)$, the chirp mass distribution $F(\M)$ also has a log-normal shape. A log-normal distribution features that $\M^xF(\M)$ also has a log-normal shape after the proper redefinition of the parameter $M_c\to \tilde M_c$. It can be checked that for large $\gamma$, the value of the redefined mass $\tilde M_c\sim M_c\sim M_0$.

We conclude that to within a factor  of two the different PBH formation models give the similar observed chirp-mass distributions and merging rates.

PBHs can be formed in clusters \cite{2001JETP...92..921R}.
Dynamical interactions in PBH clusters offers additional channel for the orbital energy dissipation thus increasing the merging rate of PBH binaries, and the constraints on $f_{pbh}$ obtained by assuming a homogeneous PBH space distribution can be lower limits. A recent analysis in \cite{Eroshenko:2023bbe} using the PBH formation model \cite{1997ApJ...487L.139N,2016PhRvL.117f1101S} shows that even  $f_{pbh}\sim 0.1-1$ is not excluded.

\section*{Conclusions}

We conclude that the chirp-mass distribution of LVK GWTC-3 BH+BH binaries with distinct two bumps can be explained by two different populations of BH+BH binaries: (1) the low-mass bump at $\M \sim 10 M_\odot$ due to the astrophysical BH+BH formed in the local Universe from the evolution of massive binaries (we have used, as an example, the model of double BH formation calculated in \cite{2019MNRAS.483.3288P} based on the simplest BH formation from the collapsing CO-core of massive stars and the standard common evolution  efficiency parameter 
$\alpha_{CE}=1$) and (2) the PBH binaries with log-normal mass spectrum (1) with $M_c\simeq 10 M_\odot$ and $\gamma\simeq 10$. The central mass of the PBH distribution is larger than the expected PBH mass at the QCD phase transition ($\sim 8 M_\odot$) but still can be accomodated with the mass of the cosmological horizon provided that the temperature $T_{QCD}\sim 70$~MeV, which is possible for non-zero chemical potential at QCD phase transition \cite{2002astro.ph.11346F}. Should the two-bump chirp-mass distribution of the merging binary BH persist in future enhanced statistics of sources detected in the forthcoming O4 LVK observing run, it can indicate the formation of PBHs with a narrow log-normal mass spectrum (1) around the central mass $M_c\sim 30 M_\odot$ comprising at least $f_{pbh}\sim 10^{-3}$ CDM fraction. Taking into account of the PBH clustering \cite{Eroshenko:2023bbe} will increase $f_{pbh}$.
        
\vskip\baselineskip
\textit{Acknowledgments.} We thank A.D. Dolgov for discussions. The authors acknowledge the support from the Russian Science Foundation grant 23-42-00055.

%\nocite{*}
%\bibliographystyle{pepan}
%\bibliography{GW1}

\end{document}